\documentclass[twocolumn]{article}
\ifx\pdftexversion\undefined
  \usepackage[dvips]{graphics}
\else
  \usepackage[pdftex]{graphics}
\fi

\usepackage{exscale,amsthm,epsfig}
\usepackage{dsfont,amssymb}
\usepackage{float}
\usepackage{mathrsfs}
\usepackage{algorithmic}
\usepackage{times}
\newcommand\comment[1]{}
\def\eps{\varepsilon}

\def\r{{R}}

\floatstyle{ruled}
\newfloat{algorithm}{tbp}{loa}
\floatname{algorithm}{Algorithm}

\newtheorem{theorem}{Theorem}
\newtheorem{lemma}{Lemma}

\theoremstyle{definition}
\newtheorem{example}{EXAMPLE}

\title{Medians and Beyond: New Aggregation Techniques \\ for Sensor
Networks}

\author{Nisheeth Shrivastava \and Chiranjeeb Buragohain \and Divyakant
Agrawal \and Subhash Suri \thanks{The authors were supported by NSF
grant IIS-0121562 and Army Research Organisation grant
DAAD19-03-D-0004 through the Institute for Collaborative
Biotechnologies.} \\ Dept. of Computer Science , University of
California, Santa Barbara, CA 93106 \\
\texttt{\{nisheeth,chiran,agrawal,suri\}@cs.ucsb.edu}}

\date{}
\begin{document}
\maketitle

\abstract{Wireless sensor networks offer the potential to
span and monitor large geographical areas inexpensively. Sensors,
however, have significant power constraint (battery life), making
communication very expensive.  Another important issue in the context of 
sensor-based information systems is that individual sensor
readings are inherently unreliable.  In order to address these two
aspects, sensor database systems like TinyDB and Cougar enable
in-network data aggregation to reduce the communication cost and
improve reliability.  The existing data aggregation techniques,
however, are limited to relatively simple types of queries such as
\texttt{SUM, COUNT, AVG,} and \texttt{MIN/MAX}.  In this paper we
propose a data aggregation scheme that significantly extends the class
of queries that can be answered using sensor networks. These queries
include (approximate) quantiles, such as the median, the most frequent
data values, such as the \emph{consensus} value, 
a histogram of the data distribution, as well as range
queries. In our scheme, each sensor aggregates the data it
has received from other sensors into a fixed (user specified) size
message. We provide strict theoretical guarantees on the approximation
quality of the queries in terms of the message size. We evaluate the
performance of our aggregation scheme by simulation and demonstrate
its accuracy, scalability and low resource utilization for highly
variable input data sets.}

\section{Introduction}

With the advances in hardware miniaturization and integration, it is
possible to design tiny sensor devices that combine sensing with
computation, storage, and communication. Availability of such devices
has made it possible to deploy them in a networked setting for
applications such as wildlife habitat monitoring~\cite{duckisland},
wild-fire prevention~\cite{firebug}, and environmental
monitoring~\cite{jamesreserve}. As new sensing devices are developed,
it is envisioned that sensor networks will be used in a large number
of civil and military applications.  Going beyond traditional
temperature, sound or magnetic sensors, a next generation of sensor
technology is emerging which can sense far more diverse physical
variables.  In particular, highly sensitive and selective
biological/chemical sensors are in development for rapid detection of
hazardous biological and chemical agents \cite{burl, chen}.

In order to support advanced sensing technology, it is necessary to
develop information and communication infrastructure in which such
sensors can be gainfully deployed. The MICA2 mote (available from
Crossbow Technology~\cite{xbow}) with TinyOS operating system
\cite{tinyos} developed at UC Berkeley represents a typical building
block of such an infrastructure. The key characteristic of MICA2 motes
is that it is severely limited in terms of computation capabilities,
communication bandwidth, and battery power. Another issue is the
inherent unreliability of the sensing functionality. Although as a
first order of approximation, sensor networks comprising multiple
sensor nodes can be viewed as a distributed system or a network of
computers, the limited capabilities of individual sensor nodes
necessitate a careful design of both the communication and information
infrastructure. Although hardware advances are likely to result in
reducing the footprint of such devices even more, the limitations and
unreliability will continue to remain. Numerous efforts are in
progress to build sensor networks that will be effective for a broad
range of applications~\cite{tinyos}.

Most common mode in which sensors and sensor networks are deployed is
in the context of monitoring and detection of critical events in a
physical environment. Typically, each sensor node collects data from
its physical environment and this data needs to be delivered to the
users through the network interconnection for further analysis. The
simplest way this can be accomplished is to let each sensor node
deliver its data periodically to the host computer, referred to as the
\emph{base station}, where the data can be assembled for subsequent
analysis. This approach, however, is wasteful since it results in
excessive communication. When combined with the fact that transmitting
one bit over radio is at least three orders of magnitude more
expensive in terms of energy consumption than executing a single
instruction, alternative approaches are clearly warranted. In order to
address this problem, proposals have been made to exploit the
multi-hop routing protocols in sensor networks in such a way that
messages from multiple nodes are combined en-route from the sensor
nodes to the base station~\cite{leach}.  Routing
in such a network can be visualized as a routing tree with the base
station as the root and nodes sending messages up the tree towards the
root.  Although this approach does reduce the number of messages, it
still suffers from the problem of larger message sizes as information
passes through the routing tree from the leaf nodes to the root node,
i.e., the base station.

Researchers at UC Berkeley~\cite{tinydb,tinydb-tag} (TinyDB project)
and Cornell University~\cite{cougar} (Cougar) have developed energy
efficient query processing architectures over sensor networks. Their
approach is based on a couple of observations : first, for a user, the
individual sensor values do not hold much value.  For example, in a
sensor network spanning thousands of nodes, the user would like to
know the average temperature of an extended region which might span a
large number of sensors.  Second, extracting all the data out of a
sensor network is very inefficient in terms of bandwidth and power
usage.  It is much more efficient to gather an overview of the total
range of data with aggregate measures such as \texttt{AVERAGE, SUM,
COUNT}, and \texttt{MIN/MAX}.  In addition to energy benefits,
aggregation can help us reduce the effects of error in sensor
readings.  Individual sensor readings are inherently unreliable and,
therefore, taking an average of multiple sensor values gives a more
accurate picture of the true physical data value.  Based on these
considerations the Cougar and TinyDB architectures have proposed using
{\em in-network aggregation} to compute such aggregates over the
routing tree, minimizing both the number of messages as well as the
size of the messages.
Note that measures such as \texttt{MIN} and \texttt{MAX} are not
strictly aggregate measures and are indeed singleton sensor values.
They are however easy to compute in the same data aggregation
framework.

Although aggregation measures such as \texttt{AVERAGE} and
\texttt{SUM} are sufficient in many applications, there are situations
when they may not be enough. In particular, in the context of
biological and chemical sensors, individual readings can be highly
unreliable and even a handful of outliers can introduce large errors
in single aggregate values such as \texttt{AVERAGE} and
\texttt{SUM}. For example, the electronic nose project~\cite{burl}
based on chemical sensors deploys a large sensor array for detecting
chemical agents.  The distribution of values on the array is used as a
chemical signature to classify the agent as being safe or unsafe.  In
such environments, we envision that it is important not only to
estimate single-valued aggregate measure but also estimate the
distribution of the sensor values. By having the estimate of the data
distribution available at the base station, users can pose more
complex queries and perform more sophisticated analysis by computing
median, quantiles, and consensus measures. Our goal in this paper is
to develop techniques that would enable such an estimate of data
distribution of sensor values be available at the base station in an
energy efficient manner while providing strict error guarantees.

Although measures such as \texttt{AVERAGE} and \texttt{MEDIAN} seem
very similar at first glance, the amounts of resource required to
compute them are very different.  To compute \texttt{AVERAGE}, every
node sends two integers to its parent, one representing the sum of all
data values of its children and the other is the total number of its
children \cite{tinydb-tag}.  In other words, \texttt{AVERAGE} can be
computed by using constant memory and by sending constant sized
messages.  On the other hand, to answer a \texttt{MEDIAN} query
accurately, we need to keep track of all distinct values and thus the
message size and memory required to store it grows linearly with the
size of the network.
To get around this difficulty we focus on \emph{approximation} schemes
to answer quantile and related queries.  For most sensor network
applications 100\% accuracy is not necessary and our approximation
scheme can be adapted to meet any user specified tolerance at the
expense of higher memory and bandwidth consumption.  To this end, we
introduce Quantile Digest or q-digest : a novel data structure which
provides provable guarantees on approximation error and maximum
resource consumption.  In more concrete terms, if the values returned
by the sensors are integers in the range $[1,\sigma]$, then using
q-digest we can answer quantile queries using message size $m$ within
an error of $\mathcal{O}(\log(\sigma)/m)$.  We also outline how we can
use q-digest to answer other queries such as range queries, most
frequent items and histograms.  Another notable property of q-digest
is that in addition to the theoretical worst case bound error, the
structure carries with itself an estimate of error for this
\emph{particular} query.

The organization of the rest of the paper is as follows.  In section
\ref{sec:background} we discuss the model we shall be working with and
some related work.  Section \ref{sec:quest} is devoted a to a detailed
description of q-digest and how it performs in-network data
aggregation.  In section \ref{sec:quest-query}, we shall show how one
can query q-digest to obtain quantities of interest.  Then in section
\ref{sec:results} we move on to an experimental evaluation of our
scheme under various inputs.  Finally we discuss extensions to
q-digest and outline directions for future work.

\section{Background  and Related Work}
\label{sec:background}

We consider a network of $n$ sensor devices, where all devices are
sensing in a common modality.  Without loss of generality, each
sensor's reading is assumed to be an integer value in the range $[1,
\sigma ]$, where $\sigma$ is the maximum possible value of the
signal. The network contains a special node, called base station,
which is responsible for initiating the query, and collecting the data
from the sensors.  When a query is initiated by the base station, the
sensors organize themselves in a spanning tree, rooted at the base
station, which acts as the routing tree for sensors to propagate their
signal values towards the base station.  Actually a routing tree is
not essential to our purposes; the only requirements we impose on the
routing scheme is that there be no routing loops and no duplicate
packets.  The routing tree can be used for query dissemination as
well.  In this paper, we assume that the links between sensor nodes
are reliable (no packets are lost), and focus exclusively on the data
aggregation problem.

An aggregate such as \texttt{MEDIAN} is intrinsically more difficult
to compute than \texttt{MIN}, \texttt{MAX}, or \texttt{AVERAGE}. In
fact, under the natural assumption that each sensor only forwards a
fixed amount of data, it is easy to argue that one cannot calculate
the median (or any other quantile) precisely. Imagine, for instance, a
simple situation where sensor $A$ calculates the median based on the
medians received from two other sensors $B$ and $C$. Even if $B$ and
$C$ know the exact median of their own data, there is an inherent
uncertainty in $A$'s computation: $A$ doesn't know the rank of $B$'s
median in dataset of $C$ and vice-versa. If $B$ and $C$ aggregate data
from $n$ sensors each, then $A$'s estimate of the combined median can
have error of $n/2$ in the worst case.

This argument shows that, with the in-network aggregation model, only
an approximation of the \texttt{MEDIAN}, or quantiles, is possible.
Our scheme, in fact, shows the best possible approximation quality
(asymptotically), and offers a trade-off between the message size and
the error guarantee.

\subsection{Related Work}
The problems of decentralized routing, network maintenance and data
aggregation in sensor networks have led to novel research challenges
in networking, databases, and algorithms \cite{intanagonwiwat,
estrin}.  In terms of providing database queries over sensor networks,
TinyDB \cite{tinydb} at UC Berkeley and Cougar \cite{cougar} at
Cornell University are the two major efforts.  They provide algorithms
for many interesting aggregates such as \texttt{MAX}, \texttt{MIN},
\texttt{AVERAGE}, \texttt{SUM}, \texttt{COUNT}.  For queries such as
\texttt{MEDIAN}, TinyDB does not perform any aggregation; all data is
delivered to the base station where \texttt{MEDIAN} is calculated
centrally \cite{tinydb-tag}.  Approximate aggregation schemes for more
complex queries such as contours and wavelet histograms have been
proposed for the TinyDB system \cite{tinydb-ipsn}.  These algorithms
perform fairly well in practice, but they do not provide any strict
bounds on error.  Zhao et al.  \cite{zhao} have also suggested
algorithms for constructing summaries like \texttt{MAX}, \texttt{AVG}.
The focus of their work is however more on network monitoring and
maintenance, rather than database query.  Considine
et. al. \cite{considine} have discussed how to compute \texttt{COUNT},
\texttt{SUM}, \texttt{AVERAGE} in a robust fashion in the presence of
failures such as lost and duplicate packets.  Przydatek
et. al. \cite{przydatek} have discussed secure ways to aggregate data,
but with only one aggregating node.  To our knowledge, this work is
the first to provide efficient approximate algorithm for queries like
quantiles, consensus and range.

The data streams community has also dealt with very similar problems
where queries on large amounts of data need to be answered with
limited memory.  In the data stream model, the data is not stored and
hence can be examined only once.  In sensor networks the data is
stored, but is distributed.  In the context of data streams, Greenwald
and Khanna \cite{greenwald} have proposed an efficient approximation
algorithm for computing quantiles.  Manku and Motwani \cite{manku}
have provided approximate algorithms for finding frequent items.
Since this paper was submitted, Greenwald and Khanna \cite{greenwald2}
have proposed a distributed sensor network algorithm to find 
approximate quantiles using message size $m$ within an error of
$\mathcal{O}(\log^2(n)/m)$.  The similarity between the problems that
arise in sensor networks and data streams suggest that it will be a
fruitful avenue of research to exploit the insights gathered on one
field on the other one.

\section{The Quantile Digest}\label{sec:quest}

A query processing framework for a sensor database needs to support
both single valued queries such as \texttt{AVG} as well as more
complex queries like \texttt{HISTOGRAM}.  Using the TinyDB framework,
many single valued queries can be answered accurately with minimal
resource usage.

\comment{As I said earlier, is COUNT really simple? Perhaps we can avoid 
a backlash by saying that these aggregates are "single valued". They are 
important, but in many applications, a more refined view of the data
is needed; e.g. the average pollution level in an area may be
misleading if the pollutants are really concentrated in a just a
few spots; an average bio-sensor reading may be meaningless, when
we really want predictions on which many sensors agree.}

In order to support more complex query functionality, we propose a new
summary structure, referred to as the q-digest (quantile digest),
which captures the distribution of sensor data approximately.
q-digest has several interesting properties which allow it to be used
in  different ways.
\begin{enumerate}

\item \emph{Error-Memory Trade-off} : q-digest is an adaptive query
framework in which users can decide for themselves the appropriate
message size and error trade-offs.  The error conscious user can set a
high maximum message size and achieve good accuracy.  A resource
conscious user can specify the maximum message size he/she is willing
to tolerate, and the q-digest will automatically adapt to stay within
this bound and provide the best possible error guarantees.  The
usefulness of this mode of operation is further extended by the
\emph{confidence factor} which is a part of q-digest.

\item \emph{Confidence Factor} : The theoretical worst case error
bound applies to only very specific data sets which are unlikely to
arise in practice.  In any actual query, the error is much smaller and
the q-digest structure contains within itself a measure of the maximum
error accumulated.  So \emph{any} answer provided by q-digest comes
with a strict bound of error.

\item \emph{Multiple Queries} : Once a q-digest query has been
completed the q-digest at the base station contains a host of
interesting information.  We can extract information on quantiles,
data distribution and consensus values from this structure without
further querying the sensor nodes.

\end{enumerate}



The core idea behind q-digest is that it adapts to the data
distribution and automatically groups values into variable sized
\emph{buckets} of almost equal weights.  Since q-digest is aimed at
summarizing the data distribution and to support quantile computation,
it is useful to compare it with traditional database approaches such
as histograms.  The critical difference between q-digest and a
traditional histogram is that q-digest can have overlapping buckets,
while traditional histogram buckets are disjoint.  q-digest is also
better suited towards sensor network queries.  For example, a simple
equi-width histogram technique is not suitable for determining
quantiles, because the weight of a bucket can be arbitrarily large
resulting in unbounded errors. For bounding errors in quantile
queries, the more appropriate approach would be to use an equi-depth
histogram~\cite{shapiro}. This technique, however, requires that the
data be stored in sorted order in a single location, which is not
practicable in a sensor network setting.  The overlapping buckets
gives q-digest another advantage over equi-depth histogram, in being
able to answer consensus queries (frequent values).

\comment{If I were to read this with a reviewers eye, I would read it
as follows: "you say that quest is superior BECAUSE it has overlapping
buckets" :-) So you need to better express this... }

The plan for the rest of this section is as follows.  First in section
\ref{subsec:q-digest-properties} we discuss the properties of q-digest
and then how one builds it in a single sensor (section
\ref{subsec:q-digest-creation}).  In section
\ref{subsec:q-digest-merge}, we show how q-digests from different
sensors are merged together.  In section \ref{sec:complexity} we prove
the memory and error bounds on q-digest. Finally, in Section 3.5, we
show how q-digest can be represented in a compact fashion.

\comment{I think the idea of using a "toy" topology, at least as presented,
is a bad one. First, it's not at all made clear "what aspect of the
problem is being abstracted" in this toy topology... In other words,
exactly what is "simpler" about the toy topology compared to the
general topologies? 
Second, I am not sure it helps me... In fact, it seems to
distract me from the main problem...}

\subsection{Properties of  q-digest}
\label{subsec:q-digest-properties}

\begin{figure}[tbh]
  \begin{center}
  \includegraphics[width=0.4\textwidth]{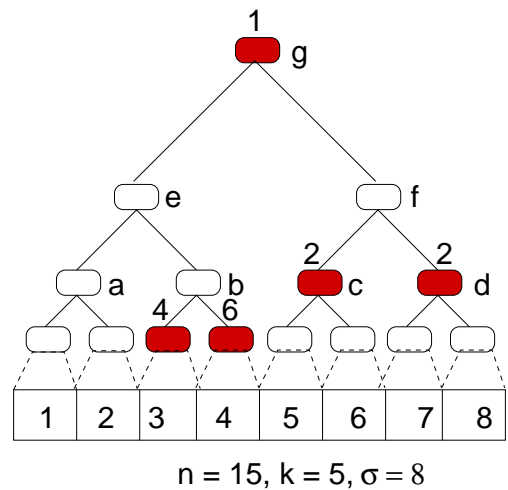}
  \end{center}
  \caption{q-digest: Complete binary tree $T$ built over the entire
    range $[1\ldots \sigma]$ of data values. The bottom most level
    represents single values. The the dark nodes are included in the
    q-digest $Q$, and number next to them represent their counts.}
\label{fig:tree}
\end{figure}

A q-digest consists of a set of buckets of different sizes and their
associated counts.  Every sensor has a separate q-digest which
reflects the summary of data available to it.  The set of possible
different buckets are chosen from a binary partition of the value
space $1, .., \sigma$ as shown in Fig.~\ref{fig:tree}.  The depth of
the tree $T$ is $\log\sigma$.  Each node $v\in T$ can be considered a
bucket, and has a range $[v.min, v.max]$ which defines the position
and width of the bucket. For example, root has a range $[1, \sigma]$,
and its two children have ranges $[1, \sigma/2]$ and $[\sigma/2+1,
\sigma]$.  The nodes at the bottom-most level have buckets of width $1$
(single values).  Every bucket or node $v$ has a counter ($count(v)$)
associated with it.

In any particular sensor, the q-digest is a subset of these possible
buckets with their associated counts.  From now on, we refer to a
q-digest as $Q$ and the \emph{conceptual} complete tree as $T$.  The
q-digest encodes information about the distribution of sensor values.
For example, the number of values which lie between $1$ and
$\sigma/2$, is the total count of all nodes in the subtree rooted at
the $[1,\sigma/2]$ node.  In Fig.~\ref{fig:tree}, the node $f$
corresponds to the range $[5\ldots 8]$ and the total number of values
in this range is $2+2=4$.  For the root node $g$ (range $[1\ldots
8]$), the total number of values is $1+2+2+4+6=15$.

The size of the q-digest is determined by a compression parameter $k$.
The exact dependence of $k$ on memory required will be spelled out in
Section~\ref{sec:complexity}.  Given the compression parameter $k$, a
node $v$ is in q-digest if and only if it satisfies the following
\emph{digest property}:

\begin{eqnarray}
count(v) & \leq & \lfloor n/k \rfloor, \label{eqn:prop1}\\
count(v)+count(v_p)+count(v_s) & > & \lfloor n/k \rfloor. \label{eqn:prop2}
\end{eqnarray}
where $v_p$ is the parent and $v_s$ is the sibling of $v$.

The only exception to this property are the root and leaf nodes. If a
leaf's frequency is larger than $\lfloor n/k \rfloor$ then too it
belongs to the q-digest. And since there are no parent and sibling for
root, its can violate property~\ref{eqn:prop2} and still belong to the
q-digest.

The first constraint (\ref{eqn:prop1}) asserts that unless it is a
leaf node, no node should have a high count.  This property will be
used later to prove error bounds on q-digest.  The second constraint
(\ref{eqn:prop2}) says that we should not have a node and its children
with low counts.  The intuition behind this property is that if two
adjacent buckets which are siblings have low counts, then we do not
want to include two separate counters for them.  We merge the children
into its parent and thus achieve a degree of compression.  This will
be described in detail in the next section.  Looking at
Fig.~\ref{fig:tree} ($n = 15, k = 5$) we can check that indeed all
nodes satisfy these two properties.

\subsection{Building a q-digest}
\label{subsec:q-digest-creation}

Consider a particular sensor $s$ that has at its disposal $n$ data
values.  Each data value is an integer in the range $[1,\sigma]$.  An
exact representation of the data will consist of the frequencies $\{
f_1,f_2,\ldots, f_{\sigma}\}$, where $f_i$ is the frequency with
which the data value $i$ is observed, and $\sum_i f_i = n$.  In the
worst case, the storage required to store this data will be
$\mathcal{O}(n)$ or $\mathcal{O}(\sigma)$, whichever is smaller.
Since transmitting this data via radio will be very expensive in a
sensor network, we would like to construct a compact representation of
this data using q-digest.  
For the ease of presentation, we shall now describe the process of
creation of a q-digest as if all the sensor data is available at $s$.
In a real sensor network all these values will be distributed across
different sensors.  We will later discuss how q-digests are
constructed in a distributed fashion on multiple sensors.

\begin{algorithm}[tbh]
\caption{COMPRESS($Q, n, k$)}
\label{compress}

\begin{algorithmic}[1]
\STATE $\ell = \log \sigma-1$;
\WHILE{$l>0$}
\FORALL{$v\ in\ level\ \ell$}
\IF{$count(v)+count(v_s)+count(v_p) < \left\lfloor \frac{n}{k} \right\rfloor$}
\STATE $count(v_p)+= count(v)+count(v_s)$;
\STATE delete $v$ and $v_s$ from $Q$;
\ENDIF
\ENDFOR
\STATE $\ell \leftarrow \ell-1 $;
\ENDWHILE

\end{algorithmic}
\end{algorithm}

To construct the q-digest we will hierarchically merge and reduce the
number of buckets.  We go through all nodes bottom up and check
if any node violates the digest property.  Since we are going bottom
up, the only constraint that can be violated is
Property~\ref{eqn:prop2}, i.e. nodes whose parent and sibling add up
to a small count. For later notational convenience we define a
relation $\Delta_v $ on the node $v$ as follows:
\begin{eqnarray*}
\Delta_v \equiv count(v) +count(v_l) + count(v_r)
\end{eqnarray*}
where, $v_l$ and $v_r$ are the left and right child of $v$.  So, if
any node $v$ whose child violate Property~\ref{eqn:prop2}, its
children are merged with it by setting its count to $\Delta_v$
and deleting its children. The algorithm to execute this hierarchical
merge is described as COMPRESS (Algorithm~\ref{compress}). It takes
the uncompressed q-digest $Q$, the number of readings $n$ and
compression parameter $k$ as input. The next example will make it
clear how the compression is done.

\begin{figure}
  \includegraphics[width=0.5\textwidth]{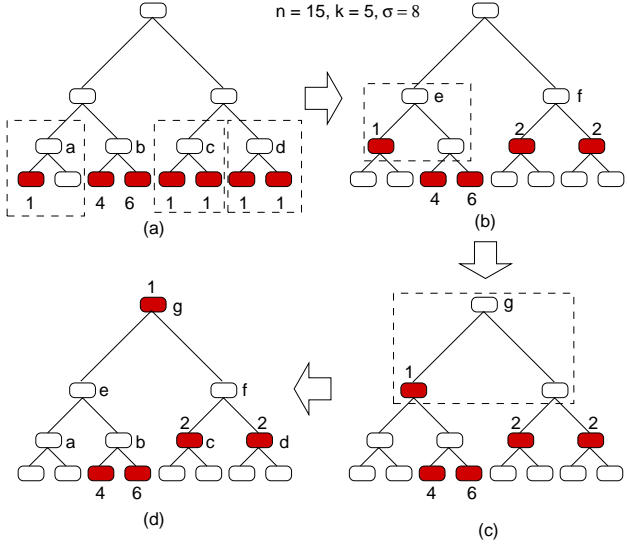}
  \caption{Building the q-digest. The leaf nodes represent values
  $[1\ldots 8]$ from left to right. Dark nodes in (d) are included in
  q-digest.}
  \label{fig:build}
\end{figure}

\begin{example}
  Consider a set of $n = 15$ values in the range $[1, 8]$ as shown in
  Fig.~\ref{fig:build}(a).  The leaf nodes from left to right
  represent the values $1, 2, \ldots, 8$ and the numbers next to the
  nodes represent the count.  The number of buckets required to store
  this information exactly is 7 (one bucket per non-zero node).  Let
  us assume a compression factor $k=5, \lfloor n/k \rfloor = 3$.  In
  Fig.~\ref{fig:build}(a), children of $a,c,d$ violate digest
  property~(\ref{eqn:prop2}).  So we compress each of these nodes by
  combining their children with them.  Thus we arrive at the situation
  in Fig.~\ref{fig:build}(b).  At this point node $e$ still violates
  the digest property.  So we compress node $e$ and arrive at
  Fig.~\ref{fig:build}(c).  Node $g$ still violates the digest
  property and so we compress $g$ and arrive at our final q-digest
  shown in Fig.~\ref{fig:build}(d).  Only $5$ nodes are required to
  store it.  $\qed$
  
\end{example}

We note some aspects of the q-digest now.  Consider node $d$ which
represents the range $[7,8]$ in Fig.~\ref{fig:build}.  The only
information that we can recover from the q-digest is that there were
two values which were present in the original value distribution in
the range $[7,8]$; the original information that there was a value $7$ and
a value $8$ has been lost.  On the other hand the information on the
ranges $3$ and $4$ have been preserved perfectly.  The q-digest can tell
us that there were exactly $4$ occurrences of the value $3$ and $6$
occurrences of value $4$.  This emphasizes a key feature of q-digest:
detailed information concerning data values which occur frequently are
preserved in the digest, while less frequently occurring values are
lumped into larger buckets resulting in information loss.

\subsection{Merging  q-digests}
\label{subsec:q-digest-merge}

\comment{

1. A skeptic will look at this and say that "Oh, they only
	know how to do toy topologies... and they are now trying to
	adapt it to realistic topologies... First of all what's
	realistic??? This works for ANY topology....

2. I suggest the following organization for the algorithm.

        Having defined the q-digest for a static data, we now
        discuss how it is constructed and propagated in the
        network. We have a spanning routing tree. The construction
        begins at the leaf nodes, and builds upward. At a sensor node
        s, the sensor s will collect the q-digests of all its
        children, plus its own (singleton value), and construct
        the q-digest for the union of these values.
        In general, the algorithm will compose multiple
        q-digests, but it is sufficient to describe the procedure
        for merging two such digests...

	So, rather than doing the insert stuff, you should explain
	the algorithm at the level of merge between two q-digests..

3. The rest I will discuss at the meeting.. I am running out of time
	and typing energy...
	
}

So far we have shown how the q-digest is built if all the data is
available on a single sensor.  But in a true sensor network setting we
need to be able to build the q-digest in a distributed fashion.  For
example if two sensors $s_1$ and $s_2$ send their q-digests to their
parent sensor (parent in the routing tree), the parent sensor needs to
merge these two q-digests to construct a new q-digest and also add its
own value to the q-digest.  A single value can be considered a trivial
q-digest with one leaf node.  Since merging multiple q-digests is no
harder than merging two digests, we shall now show how two q-digests
can be merged.

\begin{algorithm}[hbt]
\caption{MERGE($Q_1(n_1, k), Q_2(n_2, k)$)}
\label{aggr}
\begin{algorithmic}[1]

\STATE $Q \leftarrow Q_1\cup Q_2$;
\STATE COMPRESS($Q, n_1+n_2, k$);

\end{algorithmic}
\end{algorithm}

The idea is to take the union of the two q-digest and add the counts of
buckets with the same range ($[min, max]$).  Then, we \emph{compress}
the resulting q-digest. The formal MERGE algorithm is described
in Algorithm \ref{aggr}.  The following example shows the merger
of two q-digests.

\begin{figure}[htb]
  \includegraphics[width=0.5\textwidth]{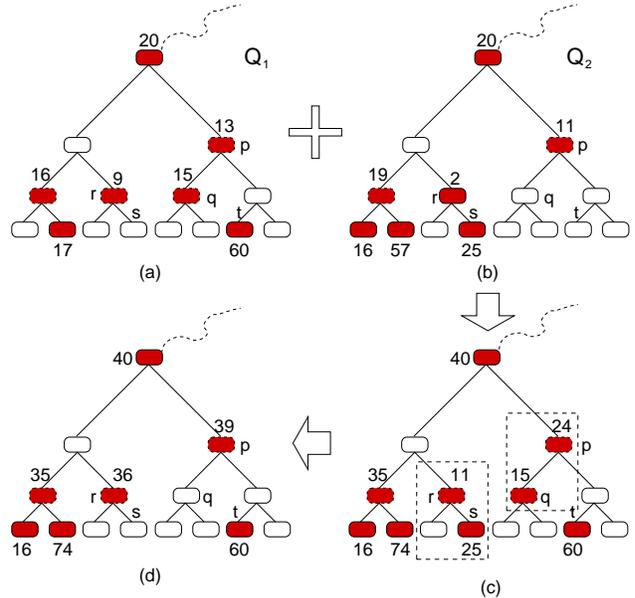}
  \caption{Merging two q-digest $Q_1$ and $Q_2$, shown in (a) and
  (b). (c) shows the union of the two q-digests. (d) is the final
  q-digest after compression.}
\label{fig:merge}
\end{figure}


\begin{example}
  Figure~\ref{fig:merge} shows the steps of merging two q-digests
  $Q_1$ and $Q_2$. For this example, $n_1 = n_2 = 200, k=10$ and
  $\sigma=64$. The tree on the left (\ref{fig:merge}(a)) shows a
  portion of $Q_1$, and tree in \ref{fig:merge}(b) shows the
  corresponding portion of $Q_2$. For the sake of clarity, we are only
  showing a small subset (range $[1\ldots 8]$) of the complete
  trees. The dark nodes are the nodes included in the q-digest,
  whereas the light ones are just for visualization.  For the final
  q-digest, $n=n_1+n_2=400$ and $\lfloor\frac{n}{k}\rfloor=40$.

  The first step is to take the union of the two q-digests. This is shown
  in Figure~\ref{fig:merge}(c). Notice the nodes in \ref{fig:merge}(a)
  and \ref{fig:merge}(b): after union, their counts have been added in
  \ref{fig:merge}(c). After this step, the q-digest could have some nodes
  which violate the digest property. In \ref{fig:merge}(c), nodes $r$
  and $p$ violate this property ($\Delta_r=36<40$,
  $\Delta_p=39<40$). (Notice that no node can violate
  Property~(\ref{eqn:prop1})). Hence, $r$ and $p$ are merged with
  their respective children (shown by the dashed rectangle). Figure
  \ref{fig:merge}(d) shows the final q-digest. \qed
\end{example}

\subsection{Space Complexity and Error Bound}
\label{sec:complexity}
In this section we evaluate the space-accuracy trade-off inherent in
q-digest.  q-digest is a small subset of the complete tree which
contains only the nodes with significant counts. This feature of the
q-digest provides the following theoretical guarantee on the size of
$Q$.

\begin{lemma}\label{lemma:memory}
  A q-digest ($Q$) constructed with compression parameter $k$ has a
  size at most $3k$.
\end{lemma}
\begin{proof}
  Since nodes in $Q$ satisfy digest property~(\ref{eqn:prop2}), we
  have the following inequality:
  \begin{eqnarray*}
    \sum_{v\in Q} \left(count(v) + count(v_p) + count(v_s)\right) >
    |Q| \cdot \frac{n}{k}
  \end{eqnarray*}
  where $|Q|$ is the size of the q-digest $Q$.

  Now, in the summation on the left hand side, the count of any node
  contributes at most once as each parent, sibling and itself. Hence,
  \begin{eqnarray*}
    \sum_{v\in Q} \left(count(v) + count(v_p) + count(v_s)\right) \\
    \leq 3 \sum_{v\in Q} count(v) = 3n.
  \end{eqnarray*}

  Hence, we get
  \begin{eqnarray*}
    |Q| \cdot \frac{n}{k}< 3n.
  \end{eqnarray*}

  So the total size of the q-digest is $3k$.
\end{proof}

\comment{
\begin{lemma}\label{lemma:memory}
  If $Q$ satisfies Property 2, it has size at most $6k$.
\end{lemma}
\begin{proof}
  We first bound the number of non-leaf nodes. Let $T'$ be the set of
  all non-leaves in $T$. Since nodes in $T'$ satisfy Property~(2), for
  each $v\in T'$, we must have $\Delta_v \geq \frac{n}{k}$.  Now, the
  weight of any node contributes to at most two $\Delta$'s (one for
  itself and one for its parent).  Therefore, we have the following
  inequality:
  \begin{eqnarray*}
    |T'| \cdot \frac{n}{k} 
    & \leq & \sum_{v\in T'} \Delta_v \\
    & = & \sum_{v\in T'} \left(count(v) + count(v_l) + count(v_r)\right) \\
    & \leq & 2 \sum_{v\in T'} count(v) \leq 2n.
  \end{eqnarray*}

  Hence, number of non-leaves are at-most $2k$, and each non-leaf can
  have two leaves (a total of $4k$ leaves), so the total size of the
  tree is $6k$.
\end{proof}
}

Any time a q-digest is created, information is lost.  As is evident
from Example 1, a node with small count will be merged into its
parent, and thus its count can recursively ``float'' to its ancestor
at any level. For example, the count of leftmost leaf in
Fig.~\ref{fig:build}(a) ends up in the root of the tree in
Fig.~\ref{fig:build}(d).  Similarly merging two digests can also lead
to information loss.  For example consider the two nodes marked as $t$
in Fig~\ref{fig:merge} (a) and (b).  In the tree $Q_2$, the
information for node $t$ has been merged into $p$.  So in the final
q-digest shown in Fig.~\ref{fig:merge}(d), the node $t$, undercounts
the occurrence of that value.  Some of that count is hidden in node
$p$ and some even in the root node.  In the worst case, the count of
any node can deviate from its actual value by the sum of the counts of
its ancestors. We will use this reasoning to prove the error bounds on
quantile queries.  This bounds the maximum error in our scheme as
shown in the next lemma.

\begin{lemma}\label{lemma:error1}
  In a q-digest ($Q$) created using the compression factor $k$, the
  maximum error in count of any node is $\frac{\log \sigma}{k}\cdot
  n$.
\end{lemma}
\begin{proof}
  Any value which should be counted in $v$ can be present in one of
  the ancestors of $v$ in $T$. So the maximum error in $v$:
  \begin{eqnarray*}
    error(v) &\leq&  \sum_{x\in ancestor(v)} count(x) \\
    & \leq & \sum_{x\in ancestor(v)} \frac{n}{k}
    \ \ \ \ \ (\textrm{Property 1})\\
    & \leq & \log \sigma \cdot \frac{n}{k}
    \ \ \ \ (\textrm{height of tree is} \log \sigma)
  \end{eqnarray*}
\end{proof}
Thus the relative error $error(v)/n$ in any node's count is
$\log(\sigma)/k$.

We now prove that after merging two q-digests, we can still maintain
the same error bounds.
\begin{lemma}\label{lemma:error}
  Given $p$ q-digests $Q_1, Q_2,... Q_p$, built on $n_1, n_2,... n_p$
  values, each with maximum relative error of $\frac{\log \sigma}{k}$,
  the algorithm MERGE combines them into a q-digest for $\sum n_i$
  values, with the same relative error.
\end{lemma}
\begin{proof}
  Merging is a two step process: union step and compression step.
  From Lemma~\ref{lemma:error1}, the compression algorithm ensures
  that the error is less than $\frac{\log \sigma}{k}$, given that the tree
  before compression had the same error bounds. So, we just need to
  prove that after the union step error is not more than
  $\frac{\log \sigma}{k}$.
  
  After union, any node $v$ of $Q$ is just the union of corresponding
  nodes $v_1, v_2,... v_p$ in q-digests, the error in $v$ can be at
  most the sum of errors in counts of $v_1, v_2,... v_p$:
  \begin{eqnarray*}
    error(v) &\leq& \sum_i error(v_i) \leq \sum \frac{\log \sigma}{k} n_i \\
    & = & \frac{\log \sigma}{k} \sum n_i = \frac{\log \sigma}{k} n 
  \end{eqnarray*}
  Hence, the relative error after union step is bounded by
  $\frac{\log \sigma}{k}$.
\end{proof}

\comment{
TODO: SHOULD WE MENTION AN ALGORITHM FOR SENSOR AS FOLLOWS??:
\begin{algorithm}[hbt]
\caption{SENSOR\_QUERY()}
\label{algo:sensor}
\begin{algorithmic}

\STATE send SENSOR\_QUERY message to $k$ children;
\STATE receive replies $T_1$, $T_2$, $\dots$ $T_k$
\STATE $T\leftarrow$ AGGREGATE($T_1$, $T_2$, $\dots$ $T_k$);
\STATE send $T$ to \emph{parent}
\end{algorithmic}
\end{algorithm}
}

Now, we prove the error bounds on quantile queries. But before we
proceed, we would like to provide a definition of quantile query and explain
how quantiles can be computed using q-digest.

In quantile query, the aim is the following: given a fraction $q\in
(0,1)$, find the value whose rank in sorted sequence of the $n$ values
is $qn$. \texttt{MEDIAN} is a special case of quantile query, with
$q=0.5$.  The relative error $\eps$ in the query is defined as
follows: if the returned value has true rank $r$, then
the error $\eps$ is
\begin{eqnarray*}
  \eps \equiv \frac{|r-qn|}{n}.
\end{eqnarray*}

We now describe how quantile queries can be answered using
q-digest. The intuition is as follow: Suppose we did a
\emph{post-order} traversal on $Q$, and summed the counts of all the
nodes visited before a node $v$. This sum $c$, is a lower bound on the
number of values which are surely less than $v.max$. We report the
value $v.max$ as $q$th quantile, for which $c$ becomes greater than
(or equal to) $qn$. This sum would be the exact quantile, if all the
non-leaf nodes whose range contains of $v.max$ (ancestors of the leaf
node containing the single value $v.max$) had a count of zero. But if
they are non zero, some of the values counted in them can be greater
than $v.max$, and we have no way to determine that. For example, if we
did a \texttt{MEDIAN} query on Fig. 2(d), we will report the value $4$
as the answer, but do not know whether the values in $g$ were less
than or more than $4$.

Using Lemma ~\ref{lemma:error}, we know that this error is bounded by
($\frac{\log\sigma}{k}\cdot n$). Hence we can find the number of
values less than $v.max$ with bounded error.  The algorithm to do this
query efficiently on a q-digest is described in
Section~\ref{sec:quest-query}.

Now we are ready to state the main result of this paper.

\begin{theorem}\label{thm:memory}
Given memory $m$ to build a q-digest, it is possible to answer any
quantile query with error $\eps$ such that
\begin{eqnarray*}
  \eps \leq \frac{3\log\sigma}{m}
\end{eqnarray*}
\end{theorem}
\begin{proof}
Choose the compression factor $k$ to be
$m/3$. Lemma~\ref{lemma:memory} says that the memory required is
$m$. The error in quantile query:
\begin{eqnarray*}
  \eps \leq \frac{\log\sigma}{k} = \frac{3\log\sigma}{m}
\end{eqnarray*}
\end{proof}

\subsection{Representation of a q-digest}

After computing the q-digest structure, each sensor has to pack it,
and transmit it to its parent. The main limitation of sensor networks
is their limited bandwidth.  To represent a q-digest tree in a
compact fashion we number the nodes from $1$ to $2\sigma-1$ in a level
by level order, i.e. root is numbered 1 and its two children are
numbered 2 and 3 etc.  Now to transmit the q-digest we send a set of
tuple of the following form $\langle nodeid(v), count(v)\rangle$ which
requires a total of $(\log(2\sigma) + \log n$) bits for each tuple.
For example, the q-digest in Fig.~\ref{fig:tree} is represented
as: $\{\langle 1, 1 \rangle, \langle 6, 2 \rangle, \langle 7, 2
\rangle, \langle 10, 4 \rangle, \langle 11, 6 \rangle\}$

\comment{TODO: building q-digest under updates....}

\section{Queries on q-digest}
\label{sec:quest-query}
In this section, we describe the possible queries that can be
supported using q-digest. We assume that the size of q-digests is $m$,
which means that the relative error $\eps$ is less than
$\frac{3\log\sigma}{m}$.

\subsection{Quantile Query}
The quantile query is: Given a fraction $q\in (0,1)$, find the value
whose rank in sorted sequence of the $n$ values is $qn$.

To find the $q$th quantile from q-digest, we sort the nodes of
q-digest in increasing right endpoints ($max$ values); breaking ties
by putting smaller ranges first.  This list ($L$) gives us the
\emph{post-order} traversal of list nodes in q-digest.  Now we scan
$L$ (from the beginning) and add the counts of nodes as they are seen. For
some node $v$, this sum becomes more than $qn$, we report $v.max$ as
our estimate of the quantile.

Notice that there are at least $qn$ readings with value less than
$v.max$, hence rank of $v$ is at least $qn$. The source of error are
readings with value less than $v.max$, present in ancestors of
$v$. These will not be counted in quantile algorithm, since $v$ comes
before its ancestors in $L$. This error is bounded by $\eps n$
(Theorem \ref{thm:memory}). So, the rank of value reported by our
algorithm is between $qn$ and $(q+\eps)n$. Thus the error in our
estimate is always positive, i.e., we always give a value which has a
rank greater than (or equal to) the actual quantile.

For example, if we perform a \texttt{MEDIAN} query on q-digest $Q$
$\{\langle 1, 1 \rangle, \langle 6, 2 \rangle, \langle 7, 2 \rangle,
\langle 10, 4 \rangle, \langle 11, 6 \rangle\}$, shown in Fig. 2(d),
the sorted list $L$ will be $\{\langle 10, 4 \rangle, \langle 11, 6
\rangle, \langle 6, 2 \rangle, \langle 7, 2 \rangle, \langle 1, 1
\rangle\}$. The count at node $\langle 11, 6\rangle$ will be more than
$0.5n$ ($8$), and we will report the value $4$ as the estimated
median. The error is bounded by the count of node $g$. 

\comment{
Notice that there are at least $qn$ readings with value less than
$v.max$. There can be some readings with values less than $v.max$,
present in ancestors of $v$. These will not be counted in quantile
algorithm, since ancestors comes after $v$ in the sorted list. Lemmas
2 and 3 prove a bound on the number of these extra values.  
}

\subsection{Other Queries}

Once the q-digest is computed, it can be used to provide approximate
answers to a variety of queries. 
\begin{itemize}

\item {\bf Inverse Quantile}: Given a value $x$, determine its rank in the
sorted sequence of the input values. 

In this case, we again make the same sorted list ($L$), and traverse
it from beginning to end. We report the sum of counts of buckets $v$
for which $x>v.max$ as the rank of $x$. The reported rank is between
$rank(x)$ and $rank(x)+\eps n$, $rank(x)$ being the actual rank of
$x$.
\item {\bf Range Query}: Find the number of values in the given range $[low,
high]$. 

We simply perform two inverse quantile queries to find the ranks of $low$
and $high$, and take their difference. The maximum error for this
query is $2\eps n$
\item {\bf Consensus Query}: Given a fraction $s\in (0,1)$, find all the
values which are reported by more than $sn$ sensors. This can be
thought of finding a value on which more than certain fraction of
sensor \emph{agreed}. These values are called \emph{Frequent
items}. 

We report all the unit-width buckets whose count are more than
$(s-\eps)n$. Since the count of leaf bucket has an error of at-most
$\eps n$ (Lemma~\ref{lemma:error1}), we will find all the values with
frequency more than $sn$.  There will be a small number of false
positives; some values with count between $(s-\eps)n$ and $sn$ may
also be reported as frequent.

\end{itemize}

\subsection{The Confidence Factor}

In Theorem~\ref{thm:memory} we proved that the worst case error for a
q-digest of size $m$ is $\frac{3\log \sigma}{m}$. But this worst
case occurs for a very pathological input set, which is unlikely in
practice. Choosing the message size according to these estimates will
lead to useless transmission of large messages, when a smaller one
could have ensured the same required error bounds. 
So if the q-digest is computed by setting $m$ to a value for which it
is \emph{expected} to deliver the required error guarantees, we still
need a way to certify that those guarantees are met. For this, we
provide a way to calculate the error in each particular q-digest
structure. We call this the \emph{confidence factor}.

If we define the \emph{weight} of a path as the sum of the counts of
the nodes in the path, the weight of the path from root to any node is
equal to the sum of its ancestors. So the maximum error is present in
the path of q-digest with the maximum weight. We define the confidence
factor $\theta$ as: $\theta =$ (maximum weight of any path from root
to leaf in $Q$) / $n$.

This ensures that the error in \emph{any} quantile query is bounded by
$\theta$.  Hence, now we can find out the maximum error in any
q-digest and discard the query if it does not satisfy the required
precision. In experiments, for example, we work with $\sigma=2^{16}$
and $m=100$, the theoretical maximum error is $\frac{3\log
\sigma}{m}\approx 48\%$, but we get a confidence factor of $\approx
9\%$ for the q-digest at the base station. This leads to huge savings
in terms of transmission cost. Notice that the actual error in query
can still be much smaller than $\theta$ (in experiments the actual
error in the median was close to $2\%$).

\section{Experimental Evaluation}
\label{sec:results}
\begin{figure}
  \includegraphics[width=0.48\textwidth]{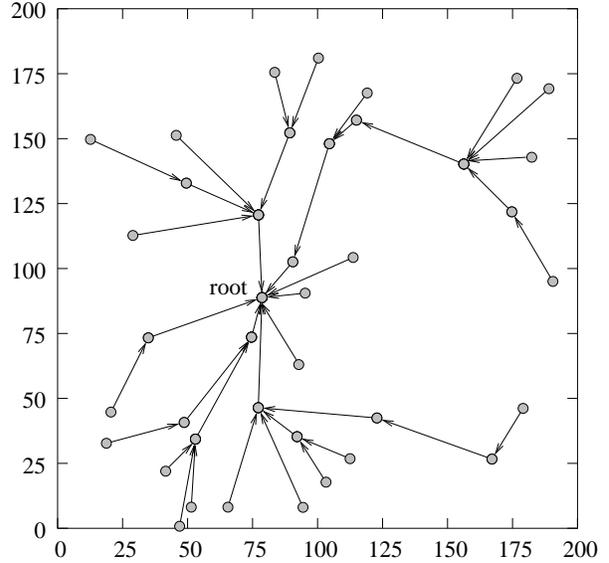}
  \caption{A typical network routing tree for 40 nodes placed in a
  200$\times$200 area.}
  \label{bfs-tree-fig}
\end{figure}
We simulated our aggregation algorithm in C++. The simulator takes the
network topology (routing tree) and readings of sensors as the
input. The base station initiates q-digest computation by sending a
query to all its children, which forwards this query to their
children, and so on. The leaf sensors send their value as q-digest to
their parent. Each sensor then aggregates q-digests received from its
children with its own reading, and then sends the aggregate to its
parent. The quantile and range queries are performed on the q-digest
received at the base station.

The topology for the network was generated as follows.  We assume that
the sensors have a fixed radio range and are placed in a square area
randomly.  If two sensors are within range of each other, they are
considered neighbors.  This generates a network connectivity graph.
The routing tree required for our simulation is simply a breadth first
search tree over this graph with an arbitrary node chosen as the root
or the base station.  In Fig \ref{bfs-tree-fig}, we show a typical
network routing tree.  When we vary the number of sensors, we vary the
size of the area over which they are distributed so as to keep the
density of sensors constant.  As an example, we used a
1000$\times$1000 area for 1000 sensors with equal radio ranges.  For
4000 sensors, the terrain dimensions were enlarged to 2000$\times$2000
keeping radio range constant.

\begin{figure}
  \includegraphics[width=0.55\textwidth]{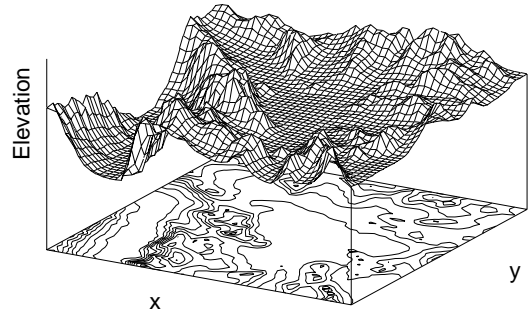}
  \caption{Three dimensional elevation data for Death Valley which is
  used to model correlated data for our simulation.  The bottom of the
  plot shows the contour lines for the terrain.}
  \label{terrain-fig}
\end{figure}

We ran our aggregation algorithm for ``random'' and ``correlated''
sensor values.  For the random case, each sensor value is taken to be
a $16$ bit random number.  In a real network, the values at sensors
are not random, but are correlated with their geographic location.  To
simulate such correlation we adapted geographic elevation data
available from the United States Geological Survey (USGS) \cite{usgs}
which is shown in Fig \ref{terrain-fig}.  The sensors are assumed to
be scattered over the terrain and the elevation of the terrain at the
sensor location is assigned as the sensor value.  The terrain size was
scaled to fit in with our simulated terrain size and the elevation
data was scaled to fit in $16$ bits.  All performance data we present
is averaged over $5$ different topologies.

We compare the performance of our algorithm with a simple unaggregated
data summarization scheme which we call \emph{list}.  In this scheme,
the summary is a list of distinct sensor values and a count for each
value. At each node, this list contains all the distinct sensor values
that occur in the subtree rooted at the node.  In other words the list
structure is a histogram with bucket width $1$.  There is no
information loss and we can answer quantile or histogram queries
exactly.  As the message progresses towards the base station, more and
more distinct values begin to occur and the size of the message grows.

\subsection{Range Queries and Histogram}
\begin{figure}[htb]
  \includegraphics[width=0.5\textwidth]{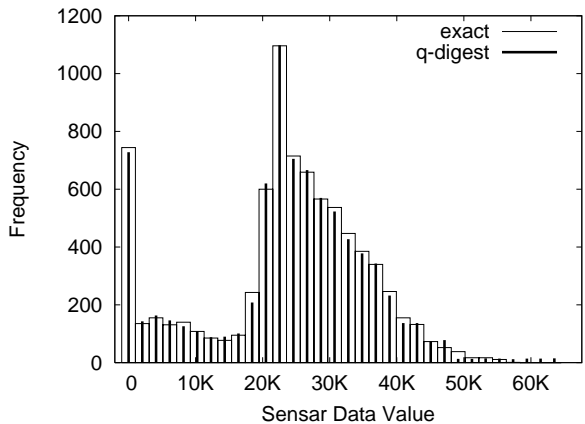}
  \caption{Exact and approximate histogram of input data shown in Fig
  \ref{terrain-fig}.  The open boxes represent the exact histogram
  while the solid thin bars represent the approximate histogram
  obtained from q-digest.}
  \label{hist-fig}
\end{figure}
As a first demonstration of our algorithm we build a histogram of the
correlated input data using range queries for 8000 nodes. We divided
the data values into $32$ equi-width buckets and queried both q-digest
and \emph{list} summaries to find the number of values in each
bucket. The resulting histogram is shown in Fig \ref{hist-fig}.  On
Fig \ref{terrain-fig} there are two relatively flat areas which are
clearly identifiable in the contour plot: the empty area near the
bottom left hand corner and the area near the center.  Sensors on
these areas will contribute a lot of values which are close to each
other.  These features lead to two peaks (at 0 and 22K) in the
histogram which are very well captured by our aggregation scheme.

\subsection{Accuracy and Message Size} 
\begin{figure}
  \includegraphics[width=0.5\textwidth]{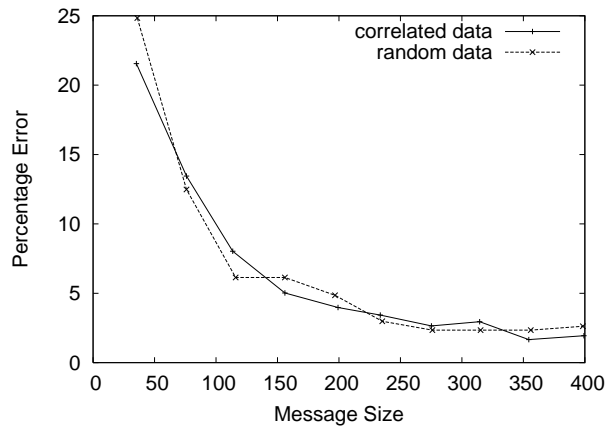}
  \caption{Measured percentage error in median vs message size (in
  bytes) for an 8000 node network.}
  \label{accuracy-fig}
\end{figure}
In an 8000 sensor network, we measured the accuracy of our algorithm
in evaluating the median for different message sizes.  The error in
this experiment is defined as the ratio of rank error in the median
estimated from q-digest and number of values ($\eps =
\frac{(|r-n/2|)}{n}$). The results are shown in Fig
\ref{accuracy-fig}.  As expected, the graph shows that the error
declines very rapidly with growing message size and with a message
size of $160$ bytes, we already are down to $5\%$ error.  There is no
significant difference in error for random or correlated data. 

We also calculated the confidence factors ($\theta$) for median
calculation with varying message sizes. This data is shown in Table
\ref{confidence-table}.
\begin{table}[bt]
\begin{tabular}{|c|c|c|c|}
\hline
  Data Type & Msg Size (bytes) & $\theta$ & Actual Error \\ \hline
  Random & $160$ & $13\%$ & $6.1\%$\\ \hline
  Correlated & $160$ & $24\%$ & $5.0 \%$ \\ \hline
  Random & $400$ & $6.6\%$ & $2.6\%$ \\ \hline
  Correlated & $400$ & $7.3\%$& $1.9\%$ \\ \hline
\end{tabular}
\caption{Maximum possible error and actual error in median query}
\label{confidence-table}
\end{table}
It is clear that the theoretically estimated accuracy is pessimistic
compared to the actual accuracy achieved.

\begin{figure}[htb]
  \includegraphics[width=0.5\textwidth]{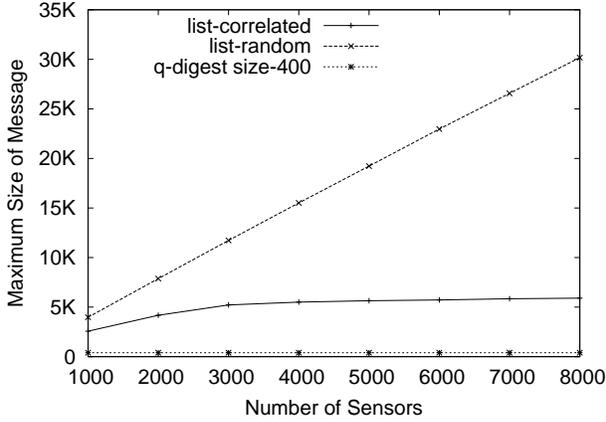}
  \caption{Maximum message sizes for different numbers of sensors for
  naive unaggregated algorithm and our aggregation algorithm.  We
  fixed message size at 400 bytes which gives about a 2\% error (see
  Fig \ref{accuracy-fig}).}
  \label{max-size-fig}
\end{figure}

Now we turn to a comparison of the message sizes required by q-digest
and those required by \emph{list}.  From Fig \ref{accuracy-fig} it is clear
that a message size of 400 is sufficient to achieve accuracy of $2\%$.
Compared to this, how much do we need to pay for exact answers?  The
comparison is shown in Fig \ref{max-size-fig} which shows maximum
message size for q-digest and \emph{list} for different numbers of sensors.
Regardless of the correlation in data values or $n$ (the number of
sensors), to achieve $2\%$ accuracy our maximum message size needs to
be no bigger than $400$.  For random data, the size for \emph{list}
increases steadily with $n$.  Since the sensor values for the random
case can be any integer between $0$ and $65535$, the number of
distinct sensor values is roughly proportional to the number of
different sensors.  For the correlated case, the number of distinct
values in the input is only about $1500$.  So the maximum message size
for \emph{list} plateaus with increasing number of sensors.

\begin{figure}
  \includegraphics[width=0.5\textwidth]{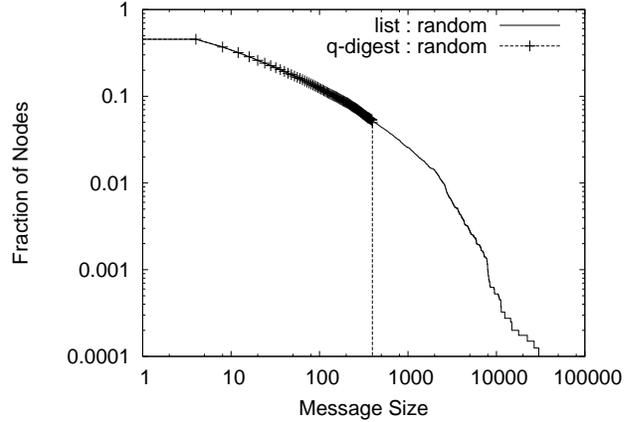}
  \caption{Cumulative Distribution of number of nodes as a function of
  message size.  On the horizontal axis we have message size $m$,
  while on the vertical axis we have number of nodes which transmitted
  messages of size larger than $m$.  Total number of nodes is 8000.}
  \label{n-vs-m-fig}
\end{figure}
A more detailed view of the distribution of message sizes is shown in
Fig \ref{n-vs-m-fig}.  Given a message size $m$, we ask the question :
what fraction of total nodes transmitted messages of size larger than
$m$?  This quantity is plotted in the vertical axis.  We compare this
distribution for $list$ and q-digest (size 400 bytes) for random input
values.  For message sizes less than 400 bytes, the \emph{list} and
q-digest the distribution is identical.  For q-digest there are no
nodes which transmit message of size larger than 400 bytes.  In
comparison, about 5\% (400 nodes) of nodes for the \emph{list} scheme
do transmit messages larger than this.  5\% might look like a small
number, but we immediately realize that these nodes actually bear an
unusually heavy load.  1\% of nodes transmit messages of size bigger
than 3K and some nodes transmit messages of size up to 30K!  These
nodes represent nodes closer to the base station.  In any routing tree
most of the nodes are near the leaf levels and such nodes are very
lightly loaded compared to nodes near to the base station.  Q-digest
does a much better job at distributing load by requiring no node to
transmit more than 400 bytes.

\subsection{Total Data Transmission}

\begin{figure}
\includegraphics[width=0.5\textwidth]{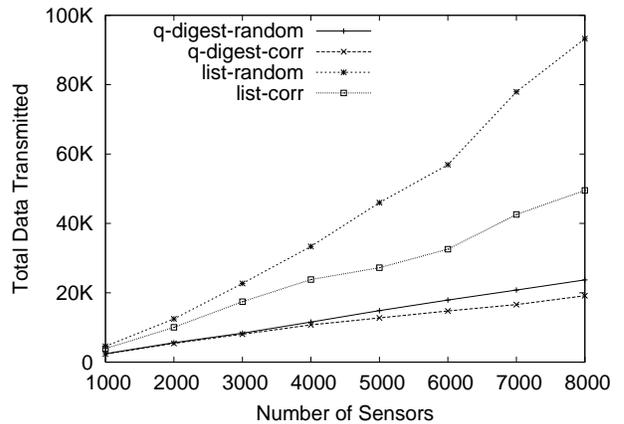}
\caption{Total data transmitted plotted as a function of total number
of sensors for both random and correlated input.  The message size for
the aggregated scheme was set at 160 bytes.}
  \label{xmission-fig}
\end{figure}

In Fig \ref{xmission-fig} we show the total amount of data transferred
for q-digest and \emph{list}.  As expected, since the number of
distinct values is less for correlated scenario, the amount of data
transferred is lower for correlated data. For a network size of
$1000$, our scheme outperforms the \emph{list} algorithm by a factor
of $2$, while for network size of $8000$, this factor increases to
about $4$. This shows that our scheme is highly scalable, and has
significant performance benefits in the case of larger networks.


\subsection{Residual Power}

Data transmission is very closely tied up with power consumption in
sensor networks.  There are two common metrics for measuring power
consumption which we shall consider in turn.

\begin{itemize}
\item \emph{Total power consumption} : This is the total power spent
by all nodes in the network and is roughly proportional to total
amount of data transmitted in the network (Fig \ref{xmission-fig}).
In reality, power consumption increases super-linearly with total data
transmitted.  This is because with increasing number of data packets,
there is more contention for the wireless medium and a lot of power
can be spent in packet collisions.
\item \emph{Lifetime} : A more appropriate power consumption metric is
the lifetime of the network.  This is the time at which network
partition occurs because of nodes running out of power.  A slightly
different definition of lifetime can be taken as the time required for
the first node to run out of power.  For a network which is geared
towards data aggregation, the nodes near the base station shoulder the
bulk of data transmission and hence runs out of power fastest.  Thus
in general, lifetime is a more useful indicator of the usable life of
the network than total power consumption.
\end{itemize}

\begin{figure}
  \includegraphics[width=0.5\textwidth]{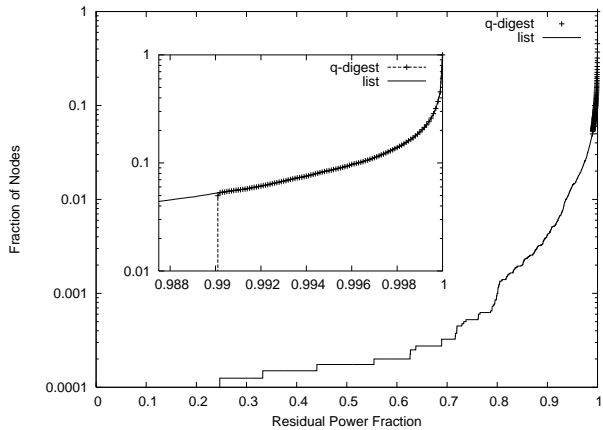}
  \caption{Cumulative distribution of nodes with residual power
  fraction.  The inset shows a magnified view of the right hand edge
  of the graph.  The total number of nodes is 8000, q-digest message
  size is 400.}
  \label{residual-power-fig}
\end{figure}

With q-digest, even nodes close to the base station transmit very
small amounts of data and the transmission burden is distributed much
more equitably.  So we can expect the usable life time of the network
to be vastly extended with our data aggregation scheme compared to the
\emph{list} scheme.  We experimentally demonstrate this by considering
the \emph{residual power} of sensor nodes after a query.  Let us
assume that all nodes in the network start with the same amount of
battery power.  After a query has been processed, different nodes will
have different amounts of power left depending on how much data each
node transmitted.  This power left is known as residual power.
Residual power is a measure of the load distribution in the network.


We simulated the effect of a single query on an 8000 node network
where all nodes started out with equal power of 40000 units.  We
assumed that for every byte transmitted, one unit of power is
depleted.  The results are shown in Fig \ref{residual-power-fig}.  On
the horizontal axis we plot residual power fraction $P$ which is
defined as
\begin{displaymath}
P =\frac{\textrm{Residual Power}}{\textrm{Initial Power}}  
\end{displaymath}
On the vertical axis we plot the number of nodes which have residual
power fraction less than $P$.  From Fig \ref{residual-power-fig} we
see that $list$ does a very bad job of distributing load.  More than
one node (0.02\% of 8000) have residual power fraction less than
$1/2$, i.e. one query drained half the battery power available for
these nodes!  At this consumption rate, after two queries using
\emph{list}, there will be at least one exhausted node.  On the other
hand q-digest performs well.  The maximum message size for q-digest
was set to 400; hence no node spent any more than 400 units of power.
Thus all nodes had residual power fraction better than 99\%.  In the
worst case, q-digest will be able to perform 100 queries before any
node runs out of power.

\section{Discussion  and Future Work}
We have presented q-digest : a distributed data summarization
technique for approximate queries using limited memory.  It accurately
preserves information about high frequency values while compressing
information about low frequency ones.  As such, it is a good
approximation scheme when there are wide variations in frequencies of
different values.  Our experimental results indicate that orders of
magnitude savings in bandwidth and power can be realized by q-digest
compared to naive schemes for both random and correlated data.  We
note that q-digest is easily extensible to multidimensional data.  For
example to handle two dimensional data, we need to extend the binary
tree representation of q-digest to a quad tree.

We have shown how a q-digest can be computed in a distributed fashion
once a query is made.  In a continuous query setting, such a digest
will become outdated as sensor values change.  It is possible to build
a new q-digest by sending in a new query; but a more efficient way
would be to send small updates such that the old q-digest can be
refreshed with new information.  

In the current work, we have not taken into account the effect of lost
messages.  The effect of lost messages can be mitigated to some extent
in a continuous query setting where the digest is continuously
updated.  In that case the parent can cache the q-digests received
from its children and if a q-digest from a child is lost, it can
replace that q-digest by the older one.

As presented in this paper, q-digest provides information about the
distribution of data values, but not information concerning
\emph{where} those values occurred.  Since q-digest is easily
extensible to multidimensional data, we are currently working on a
multi-dimensional q-digest where spatial information will be preserved
and hence the user would be able to query not only about data values,
but the spatial locations of those values as well.  We envision that
as querying architectures for sensor network become more and more
sophisticated, the use of efficient approximate algorithms will become
very common.


\end{document}